\documentclass{emulateapj}
\usepackage{epsfig}
\usepackage{graphicx}
\usepackage{float}
\usepackage{amsmath}
\usepackage{color}
\usepackage{amssymb}
\usepackage{amsfonts}
\usepackage{units}
\usepackage[colorlinks,linkcolor=blue,anchorcolor=green,citecolor=blue]{hyperref}

\shorttitle{Synchrotron heating by a fast radio burst}
\shortauthors{Yang, Zhang \& Dai}
\begin{document}

\title{Synchrotron heating by a fast radio burst in a self-absorbed synchrotron nebula and its observational signature}

\author{Yuan-Pei Yang\altaffilmark{1,2,3}, Bing Zhang\altaffilmark{3,4,5}, and Zi-Gao Dai\altaffilmark{1,2}}

\affil{$^1$School of Astronomy and Space Science, Nanjing University, Nanjing 210093, China; \\%yypspore@gmail.com;\\
$^2$ Key Laboratory of Modern Astronomy and Astrophysics (Nanjing University), Ministry of Education, China; \\
$^3$ Department of Physics and Astronomy, University of Nevada, Las Vegas, NV 89154, USA; zhang@physics.unlv.edu; \\
$^4$ Department of Astronomy, School of Physics, Peking University, Beijing 100871, China; \\
$^5$ Kavli Institute of Astronomy and Astrophysics, Peking University, Beijing 100871, China}

\begin{abstract}
Fast radio bursts (FRBs) are mysterious transient sources. If extragalactic, as suggested by their relative large dispersion measures, their brightness temperatures must be extremely high. Some FRB models (e.g. young pulsar model, magnetar giant flare model, or supra-massive neutron star collapse model) suggest that they may be associated with a synchrotron nebula. Here we study a synchrotron-heating process by an FRB in a self-absorbed synchrotron nebula. If the FRB frequency is below the synchrotron self-absorption frequency of the nebula, electrons in the nebula would absorb FRB photons, leading to a harder electron spectrum and enhanced self-absorbed synchrotron emission. In the meantime, the FRB flux is absorbed by the nebula electrons. We calculate the spectra of FRB-heated synchrotron nebulae, and show that the nebula spectra would show a significant hump in several decades near the self-absorption frequency. Identifying such a spectral feature would reveal an embedded FRB in a synchrotron nebula.
\end{abstract}

\section{Introduction}

Fast radio bursts (FRBs) are mysterious radio transients characterized by short intrinsic durations ($\lesssim$ a few $\unit{ms}$), large dispersion measures (${\rm DM}=100-2000\,\unit{pc/cm^3}$, and high Galactic latitudes ($|b|>40^\circ$) \citep{lor07,tho13}.
Their physical origin is not identified, but the observational properties place them at distances of at least extragalactic, or even cosmological \citep[e.g.][]{tho13,den14,kul14,gao14,zho14,zhe14}. Assuming such large distances, their fluences of $\sim 1\,\unit{Jy-ms}$ or larger
%, except for Lorimer FRB with a larger fluence $150 \,\unit{Jy-ms}$. Thus, the FRBs have
suggest an extremely high brightness temperature up to $\sim 10^{37}\,\unit{K}$ \citep{kat14,lua14}.

The physical origin of FRBs is unknown. Suggested models include collapses of supra-massive neutron stars to black holes \citep{fal14,zha14}, magnetar giant flares \citep{pop07,tho13,kul14}, super-giant pulses from young pulsars \citep{cor15,con15,pen15}, flaring stars \citep{loe14}, double neutron star mergers \citep{tot13}, double white dwarf mergers \citep{kas13}, evaporation of mini black holes \citep{kea12}, accretion of a comet by a neutron star \citep{gen15}, and so on. Precise localization of FRBs through radio interferometry or by detecting counterparts in other wavelengths hold the key to make the breakthrough.

In some of these proposed scenarios, FRBs are expected to be located in a surrounding nebula \citep{lyu14}. According to \cite{cor15}, \cite{con15}, and \cite{pen15}, FRBs may be super-giant pulses from young pulsars, which are probably still within their associated supernova remnants. Magnetars that produce giant flares are usually relatively young, some of which may be associated with supernova remnants \citep{mer15}. Within the ``blitzar'' scenario, \cite{fal14} suggested that delay between the birth and collapse of the supra-massive neutron stars can be thousands to even millions of years, so that one may not expect a remnant near the FRB site. However, \cite{zha14} suggested that the delay time can be much shorter, even thousands of seconds after the formation, so that a small fraction of these FRBs may be associated with GRBs, and the FRBs may be surrounded by a bright ``nebula'', i.e. radio afterglow of a GRB. Considering a continuous distribution of the delay time scale, it is possible that FRBs may be generated inside a synchrotron nebula with various ages. The existence of a ($W\propto\nu^{-4}$) scattering tail of many FRBs \citep{lor07,tho13,cha15} as well as Faraday rotation in FRB 110523 \citep{mas15} also suggest that there may exist a dense magnetized plasma screen in the vicinity of at least some FRBs \citep{wil72,lee75,ric77,mac13}.

Whereas the possible existence of a nebula in the vicinity of FRBs apparently did not affect the observed FRBs, in this paper, we consider a hypothesized scenario that some FRBs may have interacted with their nebulae, if the synchrotron self-absorption frequencies of those nebulae are above the characteristic frequency of the FRBs. This process may be named as ``synchrotron heating'' (for the nebula) or ``synchrotron external absorption'' (for the FRB). Synchrotron self-absorption (SSA) has been extensively studies by many authors \citep[e.g.][]{ree67,sar01,gao13}. Due to the contribution from the reabsorption of SSA, the electron distribution would deviate from the original (broken) power-law distribution and approach a quasi-Maxwellian distribution in the low-energy regime for a steady equilibrium situation \citep[e.g.][]{mcc69,ghi88,koo89,ghi98}.
The cross section of synchrotron absorption was calculated by \citet{ghi91}.

In this paper, we study synchrotron heating/external absorption physics within the context of FRB/nebula interaction. The theoretical framework is laid out in section 2.  The spectra is presented in section 3. The results are summarized in section 4 with some discussion.

\section{Synchrotron heating/external absorption}

Before the FRB injection, the electron distribution in the nebula is likely steady due to the balance between synchrotron energy loss and electron injection. We assume that the initial differential electron number density in the nebula is a power law:
\begin{eqnarray}
N(\gamma,0)=K\gamma^{-p},
\end{eqnarray}
where $\gamma$ is the electron Lorentz factor, $K$ and $p$ are constants\footnote{The electron distribution should be in principle a broken power law due to synchrotron cooling of the electrons. However, since an FRB only affects electrons with Lorentz factor $\gamma\sim (\nu_0/\nu_B)^{1/2}$, where $\nu_0$ is the characteristic frequency of the FRB, and $\nu_B=eB/(2\pi m_ec)$ is the electron cyclotron frequency in a magnetic field $B$, we only need to consider the electron distribution near $(\nu_0/\nu_B)^{1/2}$, where a single power law approximation is adequate.}. Before the FRB injection, the SSA intensity is given by \citep{ghi13}\footnote{We assume that the pitch angel of synchrotron radiation is $\pi/2$.}
\begin{eqnarray}
I_\nu
&=&\frac{2m_e}{\sqrt{3}\sqrt{\nu_B}}\nu^{5/2}(1-e^{-\tau_\nu})f_I(p),\\
f_I(p)&=&\frac{\Gamma[(3p-1)/12]\Gamma[(3p+19)/12]}{(p+1)\Gamma[(3p+2)/12]\Gamma[(3p+22)/12]},
\end{eqnarray}
where $\Gamma(x)$ is the Gamma function. The synchrotron absorption optical depth reads
\begin{equation}
\tau_\nu =\frac{e^2KR}{4 m_ec}\frac{1}{\nu_B}\left(\frac{\nu}{\nu_B}\right)^{-\frac{p+4}{2}}f_\alpha(p),
\end{equation}
where
\begin{equation}
f_\alpha(p)=3^{\frac{p+1}{2}}\Gamma[(3p+2)/12]\Gamma[(3p+22)/12],
\end{equation}
and $R$ is the size of the electron acceleration region. The self-absorption frequency $\nu_a$ is defined by $\tau_\nu=1$, i.e.
\begin{eqnarray}
\nu_a=\nu_B\left[\frac{\pi}{2}\frac{eRK}{B}f_\alpha(p)\right]^{\frac{2}{p+4}},
\end{eqnarray}

Let us now consider FRB injection. A fraction of low-energy electrons would be heated to higher energies by the FRB.  The kinetic equation of the electron distribution is given by \citep{mcc69}
\begin{eqnarray}
\frac{\partial N(\gamma,t)}{\partial t}&=&\frac{\partial}{\partial \gamma}\left[A\gamma^2 N(\gamma,t)\right]+\frac{\partial}{\partial \gamma}\left[C\gamma^2\frac{\partial}{\partial \gamma}\frac{N(\gamma,t)}{\gamma^2}\right]\nonumber\\
&+&S(\gamma,t).
\label{ke}
\end{eqnarray}
The first term on the right hand side describes the effect of the synchrotron energy loss, with
\begin{eqnarray}
A=\frac{1}{\gamma^2m_ec^2}\int P_s(\nu,\gamma)d\nu=\frac{2}{3}\frac{e^4B^2}{m^3c^5},
\end{eqnarray}
where $P_s(\nu,\gamma)=(\sqrt{3}e^3B/m_ec^2)F\left(\nu/\nu_{ch}\right)$ is the synchrotron power of a single electron, $F(x)=x\int_x^\infty K_{\frac{5}{3}}(\xi) d\xi$ is the synchrotron function, and $\nu_{ch}\equiv(3/2)\gamma^2\nu_B$ is the synchrotron characteristic frequency. The second term on the right hand side describes induced emission and reabsorption, with
\begin{eqnarray}
C=\frac{1}{m_ec^2}\int \frac{I_{\nu,tot}}{2m_e\nu^2}P_s(\nu,\gamma)d\nu,
\end{eqnarray}
where $I_{\nu,tot}$ is the total intensity.
The last term $S(\gamma,t)$ represents the electron injection source term\footnote{We assume that the escape of electrons from the emission region can be neglected.}. Equation (\ref{ke}) is usually used to calculate SSA \citep{mcc69,ghi88,koo89,ghi98}, but it can be also used to treat the synchrotron external absorption process discussed in this paper, since the same micro-physical processs of radiative transfer is involved \citep{mcc69}.

Compared with the broad-band synchrotron spectrum of the nebula, the FRB spectrum is relatively narrow. Observations show that the spectra of FRBs are quite steep \citep[e.g.][]{tho13,cha15}. Theoretical models invoking bunching coherent mechanisms also predict relatively narrow spectra with $\delta\nu/\nu\sim0.1$ \citep[e.g.][]{kat14}.
For the purpose of easy calculations, we approximate the FRB spectrum as a $\delta$-function, i.e. $I_{\nu,0}=I_0\delta(\nu-\nu_0)$, where $\nu_0$ is the characteristic frequency of the FRB, and $I_0$ is the effective integral intensity, which is much larger than the integral synchrotron intensity $\sim\nu_aI_\nu(\nu_a)$ of the nebula. Since $I_0\gg\nu_aI_\nu(\nu_a)$, one has $I_{\nu,tot}\simeq I_{\nu,0}$, so that the coefficient $C$ can be approximated as
\begin{eqnarray}
C=\frac{\sqrt{3}I_0 e^3B}{2m_e^3c^4}\frac{1}{\nu_0^2}F\left(\frac{\nu_0}{\nu_{ch}}\right),
\end{eqnarray}
and the derivative of the function $C$ reads
\begin{eqnarray}
\frac{dC}{d\gamma}=-\frac{2C}{\gamma}\left(\frac{\nu_0}{\nu_{ch}}\right)\frac{F'(\nu_0/\nu_{ch})}{F(\nu_0/\nu_{ch})},
\end{eqnarray}
both are needed to solve Eq.(\ref{ke}).
Before the FRB injection, the electron distribution is steady due to the balance between the synchrotron energy loss and electron injection. Thus,
$\partial(A\gamma^2N)/\partial\gamma+S(\gamma,t)\simeq0$ is satisfied, i.e. only the second term of the right hand side of Eq.(\ref{ke}) presents. After the FRB injection, the nebula electron distribution is suddenly perturbed.
The perturbation evolution can be described by
\begin{eqnarray}
\frac{\delta N(\gamma,0)}{\delta t}&=&C(p+2)(p+1)G\left(\frac{\nu_0}{\nu_{ch}},p\right)\nonumber\\
&\times&\gamma^{-2}N(\gamma,0)
\end{eqnarray}
where
\begin{eqnarray}
G(x,p)=\frac{2}{p+1}\frac{d\ln F(x)}{d\ln x}+1.
\end{eqnarray}
Here $F(x)$ is the synchrotron spectral function, and one has $d\ln F(x)/d\ln x=-x+1/2$ when $x\gg 1$, $d\ln F(x)/d\ln x=1/3$ when $x\ll 1$, and $d\ln F(x)/d\ln x=0$ when $x=0.29$.
One can define a critical Lorentz factor $\gamma_0$, so that the electron number density $N(\gamma)$ would decrease for $\gamma<\gamma_{0}$ and increase for $\gamma>\gamma_{0}$. The reason is that the electrons with Lorentz factor $\gamma\sim(\nu_0/\nu_B)^{1/2}$ would absorb the FRB's flux and be accelerated to higher energies. One may estimate $\gamma_0$ by introducing the approximation $F(x)=1.78x^{0.297}\exp(-x)$, which has an error less than $5\%$ over the range $10^{-3.5}<x<10^{0.5}$. The function $G(x,p)$ can be then approximated as $G(x,p)=2(0.297-x)/(p+1)+1$. If $G(x,p)=0$, one has $x_{0}=0.5p+0.797$. The critical electron Lorentz factor is therefore given by
\begin{eqnarray}
\gamma_{0} \simeq \left[\frac{2}{1.5p+2.391}\frac{\nu_0}{\nu_B}\right]^{1/2}.
\end{eqnarray}

Since the power-law distribution would relax to a non-power-law distribution, we need to numerically solve Eq. (\ref{ke}), and the total intensity is given by
\begin{eqnarray}
I_{\nu,tot}=I_{\nu,0}e^{-\tau_\nu^\prime}+\frac{j'_\nu R}{\tau_\nu^\prime}(1-e^{-\tau_\nu^\prime}),
\label{Itot}
\end{eqnarray}
with the first term contributed by the FRB and the second term contributed by the nebula electrons. Here
\begin{eqnarray}
j_\nu^\prime&=&\frac{1}{4\pi}\int_{\gamma_{\min}}^{\gamma_{\max}} N(\gamma,t)P_s(\nu,\gamma)d\gamma, \\
\tau_\nu^\prime&=&\frac{R}{8\pi m_e\nu^2}\int_{\gamma_{\min}}^{\gamma_{\max}}\frac{N(\gamma,t)}{\gamma^2}\frac{d}{d\gamma}[\gamma^2 P_s(\nu,\gamma)]d\gamma,
\end{eqnarray}
$N(\gamma,t)$ is the electron distribution at time $t$, $\gamma_{\min}$ and $\gamma_{\max}$ are the minimum and maximum Lorentz factors, respectively.
One can immediately see that when $\tau'_\nu \ll 1$, Eq.(\ref{Itot}) becomes $I_{\nu,tot} \simeq I_{\nu,0} + j'_\nu R = I_{\rm \nu,FRB} + I_{\rm \nu,nebula}$, so that the FRB emission goes through the nebula without significant interaction. The synchrotron heating process is only relevant when $\tau'_\nu \gg 1$.

For $\tau'_\nu \gg 1$, the perturbed nebula electron population would eventually relax to a steady state. One may estimate the relaxation time of the electron distribution. After the FRB injection, synchrotron heating from the FRB disappears, and only SSA intensity from electrons contributes to the coefficient $C$, i.e. $I_{\rm \nu,tot}\simeq I_{\rm \nu,SSA}$. For SSA, as pointed out by \citet{mcc69}, the reheating time scale is of the same order of the synchrotron cooling time scale, i.e. $t_{cool}\sim (A\gamma)^{-1}$. On the other hand, before FRB injection, since the electron distribution is steady, the time scale for electron injection is also the same order of $t_{cool}$.
Therefore, the relaxation time depends on the time scale of SSA, synchrotron cooling and electron injection, which is given by $\delta t_{\rm relax}\simeq (A\gamma_{peak})^{-1}\sim0.1[A(\nu_0/\nu_B)^{1/2}]^{-1}$, where $\gamma_{peak}$ is the peak Lorentz factor of high-energy bump in the electron distribution, and the numerical calculation shows that  $\gamma_{peak}\sim10\gamma_0$. 

We may also consider the change of the FRB spectrum. After the FRB injection, the new self-absorption frequency $\nu_a^\prime$ would be given by $\tau_\nu^\prime(\nu_a^\prime)=1$, so that $I_\nu=I_{\nu,0}e^{-\tau_\nu^\prime}$ for the FRB emission, which is significantly attenuated for $\tau'_\nu \gg 1$. In our calculation, the original FRB spectrum is a $\delta$-function. In reality, it would have a certain band width. The external absorption process becomes important below $\nu_a$, so that it essentially defines a low frequency cut-off of the FRB emission spectrum. In any case, the chance to have both an FRB and the FRB-heated nebula both observed is very low.

\section{Numerical calculation results}

With the above analysis, we numerically solve Eq.(\ref{ke}) to calculate the consequence of synchrotron heating of a nebula by an FRB. To make the problem relevant, the nebula should satisfy the following requirements: (1) The self-absorption frequency should be $\nu_a\gtrsim\nu_0$; 
%(2) The DM contributed from the nebula (assuming that the particles are not accelerated) should be much smaller than that of the FRB, i.e. $KR/(p-1)< {\rm DM} \sim 1000\,\unit{pc\,cm^{-3}}\simeq3\times10^{21}\,\unit{cm^{-2}}$; 
(2) The FRB should be close enough to the nebula so that its effective intensity is much greater than that of the nebula, i.e. $I_0 =I_{0,s}(R_s/r)^2 \gg I_s$,
where $I_s$ is the source intensity, $I_{0,s}\simeq 2\nu^2\Delta\nu kT_b/c^2\simeq1.8\times10^{27}/\Gamma\,\unit{erg\,cm^{-2}sr^{-1}s^{-1}}$ is the intrinsic intensity of the FRB, $R_s\simeq \Gamma^2 c\delta t\simeq 3\times10^7\Gamma^2\,\unit{cm}$ is the geometrical size of the FRB source, $\Gamma$ is the Lorentz factor of the FRB emitting plasma, and $r$ is the distance between the FRB source and the nebula. With these parameters, the effective integrated FRB intensity is
$I_0\simeq (1.7\times10^{15}\,\unit{erg\,cm^{-2}sr^{-1}s^{-1}})(\Gamma/100)^3(r/0.01\,\unit{pc})^{-2}$. 
For the nebula, we assume $B=1\,\unit{mG}$, which gives a self-absorption frequency 
%$\nu_a=1.4\,\unit{GHz}(B/1\,\unit{G})^{(p+2)/(p+4)}[(R/10^{13}\,\unit{cm})(K/3\times10^4\,\unit{cm^{-3}})(f_\alpha(p)/13)]^{2/(p+4)}$. 
$\nu_a=1.3\,\unit{GHz}(B/1\,\unit{mG})^{1/7}(L/10^{37}\,\unit{ergs\,s^{-1}})^{2/7}(r/0.01\,\unit{pc})^{-4/7}\times(f_I(p)/0.27)^{-2/7}$, where $L\simeq4\pi r^2[\nu_a \pi I_{\nu}(\nu_a)]$ is the nebula SSA luminosity. 
Other parameters are chosen as $R=10^{13}\unit{cm}$, $K=9\times10^{11}\,\unit{cm^{-3}}$, $p=3$, and $\delta t=1\,\unit{ms}$. 
As a result, one has $\nu_a\simeq1400\,\unit{MHz}$ so that the integrated SSA intensity of the source is $I_s\equiv\nu_aI_\nu(\nu_a)\simeq350\,\unit{erg\,cm^{-2}sr^{-1}s^{-1}}$, which is $\ll I_0$. 
%The nebula-contributed DM is ${\rm DM}_{\rm nebula} \simeq0.05\,\unit{pc\,cm^{-3}} \ll {\rm DM}_{\rm FRB}$.

The results depend on $\nu_0/\nu_a$ and $I_0/I_s$. 
We first fix the latter to $10^{15}$ and investigate the effect of the former. Given $\nu_a\simeq 1.4\,\unit{GHz}$ for our parameters, we vary the frequency of the FRB and study the electron spectrum and emission spectrum after the FRB injection as a function of $\nu_0/\nu_a$ (Fig.\ref{fig1}). One can see that as one lowers $\nu_0$, more and more nebula electrons are accelerated. 
%and the spectrum approaches Maxwellian when $\nu_0 / \nu_a \sim 0.1$ (Fig.\ref{fig1}a). 
Also the self-absorbed spectrum becomes increasingly stronger as $\nu_0/\nu_a$ decreases. The lower the $\nu_0/\nu_a$ value, the broader the modified spectral peak, since more high frequency emission is enhanced (Fig.\ref{fig1}b). It is interesting to note that the enhancement of flux is more significant at relatively higher frequencies (but not too high) above $\nu_a$. We found that  the enhancement reaches a factor of $560$ at $10 \nu_a$ for $\nu_0=0.1\nu_a$.

Next, we fix the value of the characteristic frequency of the FRB ($\nu_0=0.5\nu_a=700\,\unit{MHz}$), and change the intensity ratio $I_0/I_s$ around $10^{16}$. Similar to decreasing $\nu_0/\nu_a$, increasing $I_0/I_s$ would lead to acceleration of more electrons. 
%However, the electron distribution does not approach Maxwellian with increasing $I_0/I_s$ (Fig.\ref{fig2}a), suggesting that the asymptotic electron energy distribution shape is defined by $\nu_0/\nu_a$ rather than $I_0/I_s$. 
The self-absorbed emission spectrum also shows a more significant hump feature as $I_0/I_s$ increases (Fig.\ref{fig2}b).

%\textbf{Figure \ref{fig3} shows the predicted light curves of FRB synchrotron heating of a nebula at different frequencies and for $I_0=10^{16}I_s$ and $\nu_0=0.5\nu_a$.} Noticing that the light curve is in logarithmic scale, one can see that the heating is instantaneous. 
%\textbf{The light curve is essentially flat, lasting for a long time defined by $\delta t_{\rm relax}\sim[A(\nu_0/\nu_B)^{1/2}]^{-1}$, which is a few ten thousand years for the parameters adopted in our calculations.}
%After $\delta t_{\rm relax}$ the flux decays roughly exponentially before approaching the original nebula flux. Such a light curve feature would be regarded as an observational signature of FRB heating of a nebula. It is interesting to note that the enhancement of flux is more significant at relatively higher frequencies (but not too high) above $\nu_a$. \textbf{In the calculations (Fig.\ref{fig3}), the enhancement reaches a factor of $80$ at $10 \nu_a$.}

Finally, we consider the modification of the FRB spectra due to the absorption of the nebula. As shown in Fig. \ref{fig3}, the new nebula self-absorption frequency $\nu_a^\prime$ after the FRB injection is larger than the original self-absorption frequency $\nu_a$ before the FRB injection. The FRB is subject to such absorption. Only at the high-energy band where $\nu > \nu'_a$ is satisfied can the emission escape the source. The emission at $\nu < \nu'_a$ is absorbed and used to accelerate nebula electrons. Given a steep spectral index for most FRBs \citep{tho13,cha15}, the escaped emission would be a very small fraction of the total FRB flux, and therefore would not be detectable.

\section{Conclusions and Discussion}

Motivated by the possibility that FRBs may be physically located in a synchrotron nebula in some progenitor models \citep[e.g.][]{zha14,con15,pen15}, in this paper we investigate the physical process of synchrotron heating by the FRB in a self-absorbed synchrotron nebula. We show that if the FRB characteristic frequency $\nu_0$ is below the self-absorption frequency $\nu_a$ of the nebula, the nebula would undergo a sudden brightening, leading to a spectral hump signature near $\nu_a$. The lower the $\nu_0/\nu_a$ ratio and the larger the $I_0/I_s$ ratio, the more significant the heating signature. Identifying such a signature would support the origin of FRBs in synchrotron nebulae, such as the super-giant-pulse models for young pulsars \citep{cor15,con15,pen15}, magnetar giant flare models \citep{pop07,tho13,kul14}, or supra-massive neutron star collapse model \citep{fal14} in a relatively young nebula \citep{zha14}.

For typical parameters adopted in this paper, the relaxation time after FRB synchrotron heating is extremely long, reaching a few thousand years for $B \sim 1$ mG. Within the observational time scale, this can be regarded as a permanent distortion of the spectrum. Identifying the spectral bump feature would then give a strong evidence of FRB heating in the nebula. Interestingly, if the nebula has a stronger magnetic field (e.g. $B\sim1\,\unit{G}$), such as the case of a GRB afterglow or the afterglow of pre-FRB explosion \citep[e.g.][]{zha14}, our model would predict a type of radio transient lasting for $\delta t_{\rm relax}\simeq27\,\unit{days}(B/1\,\unit{G})^{-3/2}(\nu_0/1.4\,\unit{GHz})^{1/2}$, which is followed by an exponentially decay to the original nebula flux level due to the relaxation of electrons to the original distribution.

%The absorption and re-emission of a synchrotron nebula is dependent on its properties. As we discussed above, one of the most important parameters is the nebula self-absorption frequency $\nu_a=1.4\,\unit{GHz}(B/1\,\unit{G})^{(p+2)/(p+4)}[(R/10^{13}\,\unit{cm})(K/3\times10^4\,\unit{cm^{-3}})(f_\alpha(p)/13)]^{2/(p+4)}$, which should be larger than the FRB characteristic frequency $\nu_0$. If a FRB could be absorbed, there should exist a dense magnetized plasma in the nebula.  Such a signature could constrain the properties of the nebula around a FRB.
The fraction of FRBs that are absorbed depends on the relative $I_\nu$ ratio and, more importantly, the distribution of nebula $\nu_a$ with respect to the FRB band ($\nu_1, \nu_2$): An FRB is completely absorbed if $\nu_2 < \nu_a$, partially absorbed if $\nu_1<\nu_a<\nu_2$, and not absorbed if $\nu_1>\nu_a$.
%In fact, there will be significant fraction of the FRB emission needs to be absorbed by the nebula for both nebula and FRB to be observed. It depends on the relative $I_\nu$ ratio which depends on the distance between the FRB and the nebula and the intrinsic brightness of the FRB. On the other hand, the fraction also depends on the FRB frequency band $(\nu_1,\nu_2)$. 
%If $\nu_2<\nu_a$, almost all of the FRB emission is absorbed, we can only observe the heated nebula. If $\nu_1>\nu_a$, the FRB emission is not absorbed by nebula at all, and we can observe the FRB and nebula (both nebula and FRB emission do not change). If $\nu_1<\nu_a<\nu_2$, only part of FRB with low frequency could be absorbed by the nebula, we also can observe the FRB and heated nebula.}
 For the parameters of our calculation, a large nebula luminosity $L$ is needed to get a high enough $\nu_a$ to make our problem relevant. Observationally FRBs are in the GHz range, suggesting that these FRBs are usually not absorbed. It is possible that some FRBs may have lower peak frequencies, so that they could be absorbed by dimmer nebula. Future surveys of faint, low-frequency synchrotron nebula would constrain whether absorbed low-frequency FRBs indeed exist in nature.
%For some nebula with low-frequency $\nu_a$ some FRBs in principle may have lower peak energies. However, they are likely absorbed, so that the observed ones typically have higher frequencies in the GHz range.}
 
\begin{figure}[H]
\centering
\includegraphics[angle=0,scale=0.3]{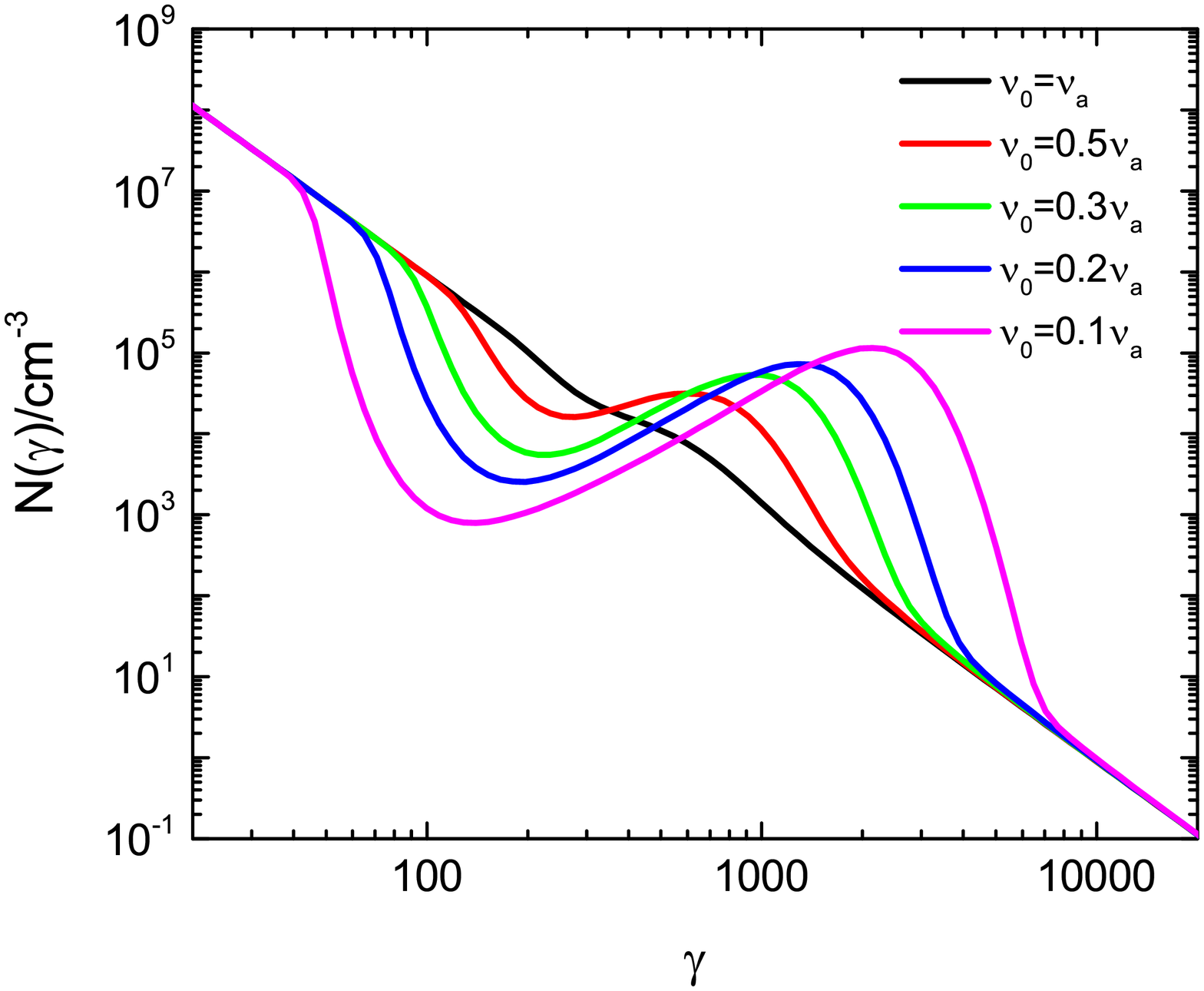}
\includegraphics[angle=0,scale=0.3]{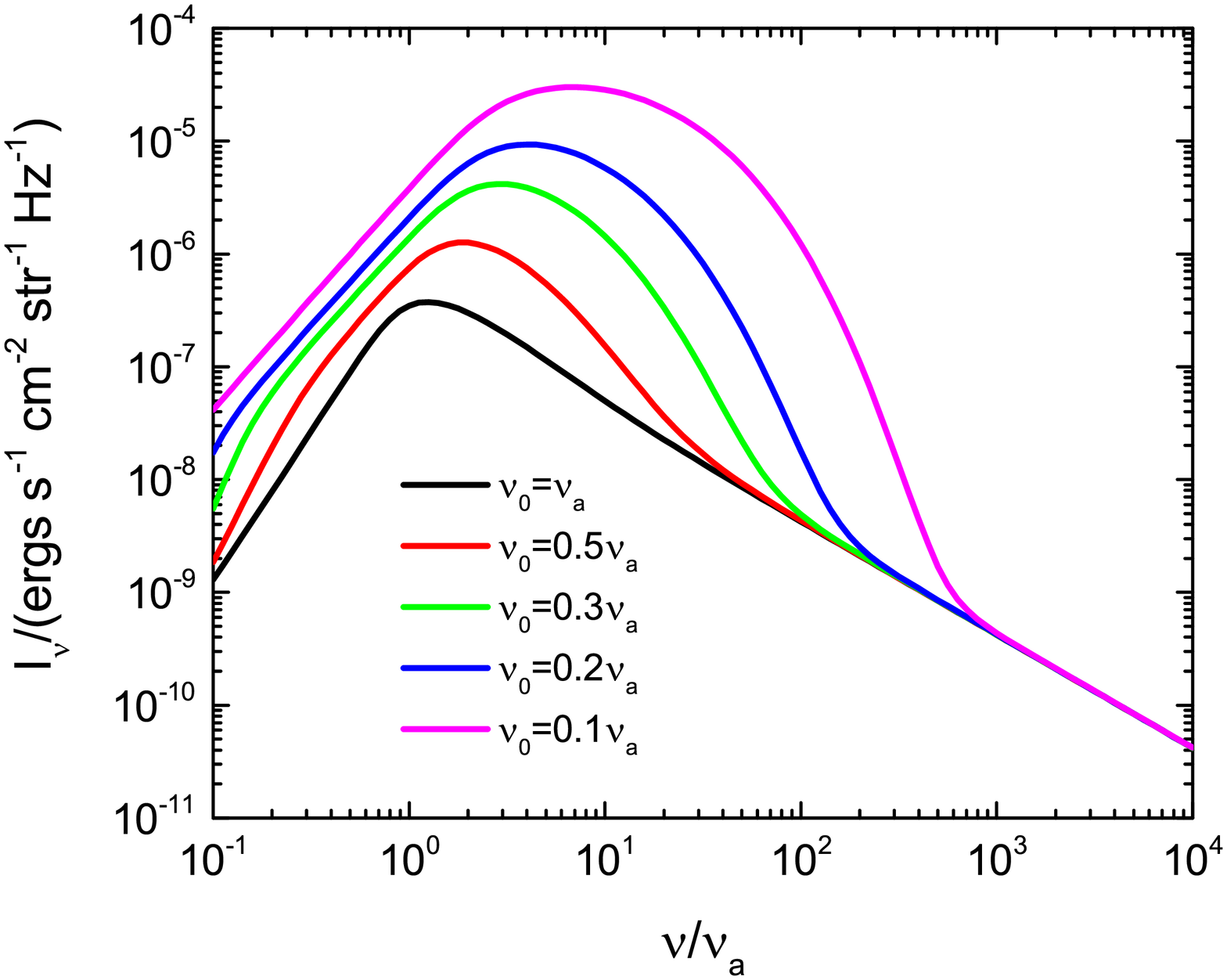}
\caption{Consequence of FRB synchrotron heating and its dependence on $\nu_0/\nu_a$ with $I_0/I_s=10^{15}$ fixed: (a) The electron spectra; (b) The emission spectra. In both cases, different colors denote different $\nu_0/\nu_a$ ratios.} \label{fig1}
\end{figure}

\begin{figure}[H]
\centering
\includegraphics[angle=0,scale=0.3]{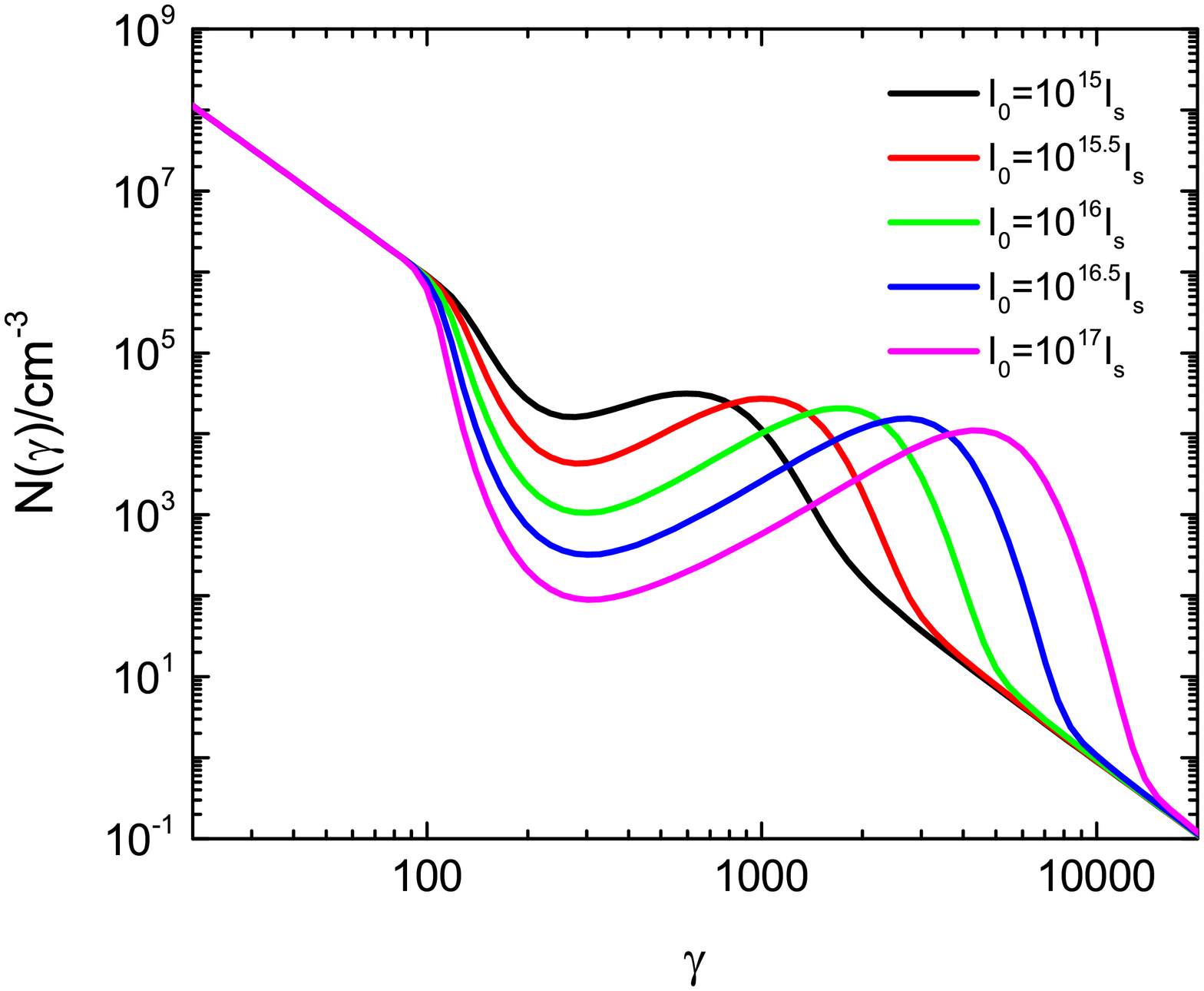}
\includegraphics[angle=0,scale=0.3]{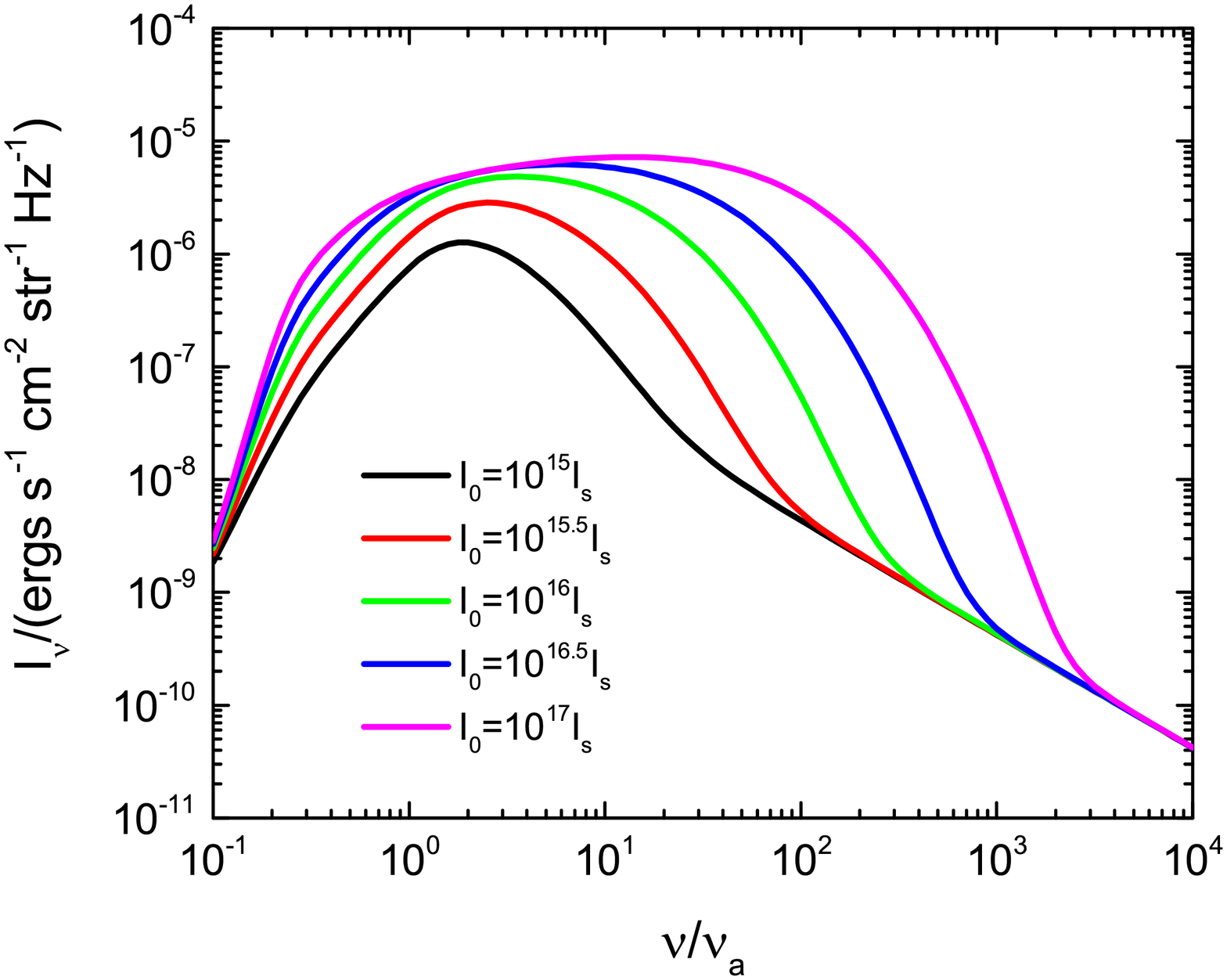}
\caption{Consequence of FRB synchrotron heating and its dependence on $I_0/I_s$ with $\nu_0/\nu_a = 0.5$ fixed: (a) The electron spectra; (b) The emission spectra. In both cases, different colors denote different $I_0/I_s$ ratios.} \label{fig2}
\end{figure}

\begin{figure}[H]
\centering
\includegraphics[angle=0,scale=0.3]{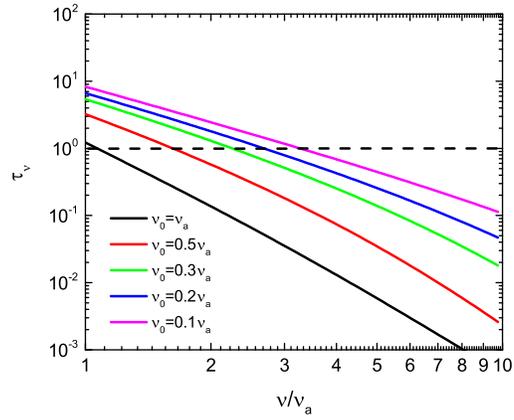}
\caption{The optical depths as a function of $\nu_0/\nu_a$ for different FRB characteristic frequency $\nu_0/\nu_a$ (different colors), where $\nu_a$ is the original SSA frequency of the nebula. The parametets $I_0/I_s = 10^{15}$ is adopted. The new self-absorption frequency $\nu'_a$ is defined by the condition of $\tau'_\nu = 1$ (dashed line), which increases with decreasing $\nu_0$.}\label{fig3}
% colored lines denote different FRB's characteristic frequencies $\nu_0$. We assume that the effective integral intensity of FRB is $I_0=10^{10}I_s$, where $I_s\equiv\nu_aI_\nu(\nu_a)$ is the integral intensity of SSA before FRB injection. $\nu_a$ is the self-absorption frequency of SSA before FRB injection.}\label{fig4}
\end{figure}

It is interesting to calculate the flux of the nebula. 
For our parameters, the predicted nebula flux before FRB injection is given by $F_\nu=\pi I_\nu(r/d)^2=3\,\unit{\mu Jy}(I_\nu/10^{-7}\,\unit{ergs\,s^{-1} cm^{-2} str^{-1} Hz^{-1}})(r/0.01\,\unit{pc})^2\times(d/1\,\unit{Gpc})^{-2}$, where $I_\nu$ is the nebula intensity, $r$ is the nebula radius, and $d$ is the luminosity distance. 
This flux is low, which is comparable to or even lower than the flux of the radio afterglow of an FRB itself \citep{yi14}. 
After FRB injection, for an FRB and associated nebula with $\nu_0=0.1\nu_a$ at $d\simeq1\,\unit{Gpc}$, the SSA flux from the nebula at high frequency (e.g. $\nu\simeq10\nu_a$) would be significantly increase, $F_\nu\simeq1\,\unit{mJy}$, as shown in Fig. \ref{fig1}b.
Such a nebula may be detectable 
%if the FRB source is nearby, e.g. $d\lesssim10\,\unit{Mpc}$, 
by the current or upcoming radio telescopes such as Australian Square Kilometer Array Pathfinder (ASKAP), Five-hundred-meter Aperture Spherical radio Telescope (FAST), and LOw-Frequency ARrray (LOFAR) and VLA Sky Survey (VLASS).

%Interestingly, the LOFAR team recently reported a low-frequency (60 MHz) transient near the North pole, which lasted for about 10-minute duration and is dubbed ILT J225347+862146 \citep{ste15}. The origin of such source is mysterious, and not expected from conventional models. We note that even though the observed parameters are somewhat different from the nominal parameters adopted in our paper (e.g. lower frequency and shorter duration), the general signature (flat light curve followed by a decay) is consistent with the signature predicted in this paper, and by adjusting parameters (e.g. a low-frequency FRB and a stronger $B$ field in the nebula), the model may explain this mysterious transient. A prediction of this model is that there could be a faint synchrotron nebula at the location of the transient. A deep radio observation at the source location is encouraged to test this model.

\acknowledgments
We thank the anonymous referee for valuable and detailed suggestions that have allowed us to improve this manuscript significantly. 
B.Z. acknowledges insightful discussion with Shri Kulkarni and Sterl Phinney that stimulated this research. We thank Xue-Feng Wu, Wei-Hua Lei and Long-Biao Li for helpful discussion. This work is partially supported by National Basic Research Program (973 Program) of China under grant No. 2014CB845800 and the National Natural Science Foundation of China under grant No. 11573014. Y.P.Y. was supported by China Scholarship Program to conduct research at UNLV.

%\clearpage

%The differential electrons number density as a function of their Lorentz factors for various FRB's characteristic frequencies. Differently colored lines denote different FRB's characteristic frequencies $\nu_0$. We assume that the effective integral intensity of FRB is $I_0=10^{10}I_s$, where $I_s\equiv\nu_aI_\nu(\nu_a)$ is the integral intensity of SSA before FRB injection. $\nu_a$ is the self-absorption frequency of SSA before FRB injection.}\label{fig1}
%\end{figure}

%\begin{figure}[H]
%\centering
%\includegraphics[angle=0,scale=0.4]{fig2.eps}
%\caption{The SSA spectrum of the nebula electrons for various FRB's characteristic frequencies. Differently colored lines denote different FRB's characteristic frequencies $\nu_0$. We assume that the effective integral intensity of FRB is $I_0=10^{10}I_s$, where $I_s\equiv\nu_aI_\nu(\nu_a)$ is the integral intensity of SSA before FRB injection. $\nu_a$ is the self-absorption frequency of SSA before FRB injection.}\label{fig2}
%\end{figure}

%The differential electrons number density as a function of their Lorentz factors for various FRB's effective integral intensities. Differently colored lines denote different FRB's effective integral intensities. We assume that the characteristic frequency of FRB is $\nu_0=0.5\nu_a$, where $\nu_a$ is the self-absorption frequency of SSA before FRB injection. $I_s\equiv\nu_aI_\nu(\nu_a)$ is the integral intensity of SSA before FRB injection.}\label{fig3}
%\end{figure}

%\begin{figure}[H]
%\centering
%\includegraphics[angle=0,scale=0.4]{fig4.eps}
%\caption{The SSA spectrum of the nebula electrons for various FRB's effective integral intensities. Differently colored lines denote different FRB's effective integral intensities $I_0$. We assume that the characteristic frequency of FRB is $\nu_0=0.5\nu_a$, where $\nu_a$ is the self-absorption frequency of SSA before FRB injection. $I_s\equiv\nu_aI_\nu(\nu_a)$ is the integral intensity of SSA before FRB injection.}\label{fig4}
%\end{figure}

%\begin{figure}[H]
%\centering
%\includegraphics[angle=0,scale=0.3]{fig3.eps}
%\caption{The predict transient light curves of a FRB-heated synchrotron nebula. The parametets $I_0/I_s = 10^{16}$ and $\nu_0/\nu_a = 0.5$ are adopted.}\label{fig3}
%%the SSA emission of the nebula electrons for various observed frequencies. Differently colored lines denote differently observed frequencies. We assume that the characteristic frequency of FRB is $\nu_0=0.5\nu_a$, where $\nu_a$ is the self-absorption frequency of SSA, and the FRB's effective integral intensity is $I_0=10^{11}I_s$.}\label{fig5}
%\end{figure}

\end{document}